# Natural games

**Jani Anttila [a] and Arto Annila [a,b,c,*]** ,

[a] Department of Biosciences, FI-00014 University of Helsinki, Finland
[b] Institute of Biotechnology, FI-00014 University of Helsinki, Finland
[c] Department of Physics, FI-00014 University of Helsinki, Finland

**ABSTRACT**

Behavior in the context of game theory is described as a natural process that follows the 2nd law of thermodynamics. The rate of entropy increase as the payoff function is derived from statistical physics of open systems. The thermodynamic formalism relates everything in terms of energy and describes various ways to consume free energy. This allows us to associate game theoretical models of behavior to physical reality. Ultimately behavior is viewed as a physical process where flows of energy naturally select ways to consume free energy as soon as possible. This natural process is, according to the profound thermodynamic principle, equivalent to entropy increase in the least time. However, the physical portrayal of behavior does not imply determinism. On the contrary, evolutionary equation for open systems reveals that when there are three or more degrees of freedom for behavior, the course of a game is inherently unpredictable in detail because each move affects motives of moves in the future. Eventually, when no moves are found to consume more free energy, the extensive-form game has arrived at a solution concept that satisfies the minimax theorem. The equilibrium is Lyapunov-stable against variation in behavior within strategies but will be perturbed by a new strategy that will draw even more surrounding resources to the game. Entropy as the payoff function also clarifies motives of collaboration and subjective nature of decision making.

## 1. Introduction

Game theory for the mathematical modeling of human behavior originates from John von Neumann and John Nash who both drew inspiration from the behavior of thermodynamic systems [1]. Curiously though applications of game theory in physics and chemistry remain few whereas the breadth of applications has grown impressive otherwise and covers diverse disciplines, most notably economics [2], biology [3] and social sciences [4,5] as well as engineering [6], computer and information sciences [7] and also extends to models of biochemical and biophysical processes [8]. Therefore, when considering the broad scope of game theory, could it be, just as the pioneers envisioned, that there is after all a profound connection between behavior and physical processes via a universal natural law that underlies the model of behavior in the many forms of games?

---
* Corresponding author.
 Email address: arto.annila@helsinki.fi

The common basic structure of all games implies on one hand that there is some universal principle underlying behavior of systems. On the other hand, game theory contains a considerable dispersion of variants, each tailored to model outcomes of specific strategic situations. In particular there is no consensus about a universal payoff function, denoted $\pi_i(s)$ for a strategy profile $s = (s_1,...,s_i)$, whose maximization, i.e., payoff $\pi_i(s^*)$ for the optimal strategy $s^*$, would cover all incentives of player $i$ in competition with other players $-i$. The difficulty in formulating a universal theory of games that would enclose diverse disciplines seems to relate to what exactly the behavior aims to maximize. In fact, when considering the variety of circumstances that are confronted during various decision making processes, it may appear inconceivable that there could possibly be a universal payoff function.

In economics the payoffs are customarily summed up simply as money or, when all other desirables are included, the payoff is referred to as 'utility' $u_i(s)$ or expected utility $E_i(s)$. On the other hand, utility is an elusive concept, which makes it difficult to assign realistic numerical payoff values. Thus, it seems problematic to find an accurate and quantitative formulation for all motive forces that underlie behavior. Moreover, in the light of any chosen payoff, an individual (referred to as a player) appears at times to act irrationally, e.g., seems to settle for suboptimal personal payoff depending on other players' actions, even when considered from the perspective of optimizing over multiple sequential games. Also, players may seem to act inconsistently with respect to their past actions, sometimes almost 'for the sake of it'. Various ways to explain these perplexing findings with bounded rationality [9] have been pursued. While a consensus is still lacking, the discussion may benefit from the ideas offered herein.

In biology, too, the choice of payoff is crucial to modeling behavior realistically. The payoff may be equated with fitness but similarly to utility there is no generally accepted way of quantifying fitness [10]. Depending on the situation, either reproduction rates or steady-state population densities could be used, but these may lead to different outcomes [8]. Nevertheless, in the context of behavioral ecology, an important breakthrough was made when natural selection alone as the profound principle, was found to prevent alternative strategies from invading a population that is practicing an evolutionarily stable strategy [11].

In this paper we will re-inspect the theory of games from the physical perspective using the second law of thermodynamics. The intention here is not to present a new model or to improve on game theoretical models. Instead, the aim is to map game theoretical concepts into their physical counterparts and to elucidate that game theory is successful in modeling the behavior of various kinds of systems because the behavior itself is a physical process that is governed by a universal principle. This task, despite paralleling the original ideas of von Neumann and Nash, is motivated by the recent derivation of the principle of increasing entropy as an equation of motion from the statistical physics of open systems [12]. The revised statistical theory is not limited to closed systems but describes various natural processes, both inanimate and animate, including behavior of biological, economic and cultural systems [13,14,15,16,17,18,19,20,21,22]. According to the 2nd law all these systems at various levels of nature's hierarchy consume free energy as quickly as possible. This observation leads to the proposition that the behavior of many systems, including decision-making processes, could be described as a natural process so that entropy is the universal payoff function.

## 2. The Progress of a Game as a Natural Process

The theory of games, as it was formulated by von Neumann, is based on thermodynamics which, in turn, follows from statistical mechanics. When depicting behavior as a physical process directed by various incentives, it was not strange for von Neumann, as a physicist, to compare all exchangeable entities, i.e., assets, in terms of energy and to relate any asset to the average energy density per entity $k_BT$ in the thermodynamic system. It was only when applying the thermodynamic formula to complex practical systems that von Neumann, in collaboration with economist Oskar Morgenstern abandoned the principle idea to map one-to-one a physical system to an economical system [1]. When announcing that money will serve as the payoff, von Neumann and Morgenstern were deeply aware of the limitations caused by the adopted approximation but could not do better.

Likewise, when Nash formulated the equilibrium concept that carries his name, he adopted the notion of chemical equilibrium from Gibbs [23]. However, Nash did not aim to make a one-to-one mapping of mathematical variables to energy densities of chemical compounds but recognized the resemblance. Nonetheless it is common, especially in economics [24], to see an analogy between the progress of a chemical reaction toward the thermodynamic equilibrium and the progress of a game toward a Nash equilibrium. The similarities between mathematical models and physical realizations may be even more apparent in evolutionary game theory where mixed strategies $s_i$ are commonly interpreted as portions of a population expressing a specific behavior. According to this interpretation, suggested by Nash [25], population densities keep changing, just like reactant concentrations, until a stable point is reached. The steady state of an ecosystem depends on surrounding conditions, just as the chemical equilibrium is a function of temperature according to Le Chatelier's principle [26]. The stability of a solution condition may also be lost when a new strategy emerges, corresponding to an addition of a new agent, such as a catalyst, into a chemical reaction mixture. Likewise, the stationary point may shift when a new species is introduced in an ecosystem. The new species with its characteristic behavior may use resources that were unreachable to its predecessors. At these critical events [27], known also bifurcations [28], the old, inferior strategies give way to new, superior strategies as the system evolves further.

Since von Neumann and Nash, studies in human behavior in various controlled circumstances, referred to as games, have given rise to stricter and broader solution concepts to classify equilibria as well as to different kinds of games to model various situations. For example, in the basic zero-sum game the players exchange assets among each other whereas in a non-zero-sum game they also compete for external resources. In terms of physics the former variant of the general theme corresponds to a closed and fixed energy ensemble that behaves as a Hamiltonian system [29] whereas the latter corresponds to an open system that acquires energy from its surroundings. A game is called non-generic if a small change to one of the payoffs may remove or add a Nash equilibrium. This means in terms of physics that such a system is not sufficiently statistical. For example a corresponding extensive-form game will progress further when a player with a new strategy gains access to additional assets from the exterior and therefore extending the boundaries of the game. This will increase the payoff. In the same way a non-Hamiltonian thermodynamic system will evolve further when a new reaction component or mechanism gains access to additional free energy. This path-dependent process will increase entropy which, according to the basic maxim of chemical thermodynamics is equivalent to the decrease in free energy.

In the quest for a universal theory of games it is of interest that evolutionary game theory demonstrates scale-independents forms of games [30]. It is successful in explaining various ecological scenarios where populations compete with each other in the same way as individuals. Moreover, evolutionarily stable strategies of population games and the Nash equilibria of decision-making games usually coincide. These examples of scale-free games imply that sentient, population and inanimate processes are basically alike and operate under a common imperative, only at different scales. Yet there has been much debate about what exactly the players, may they be molecules, cells, individuals or populations, aim at maximizing. It seems that a common universal payoff function is required to unite the diverse models and to place the theory of games on a profound principle. Here this possibility is examined using statistical physics of open systems.

## 3. Thermodynamic Formulation of Behavior

The theory of games as it was formulated by von Neumann and expanded by Nash is founded on Boltzmann's astounding idea that nature is in motion toward increasingly more probable states. Boltzmann adopted the probability concept from Descartes, Fermat, Pascal and others who had computed combinatorial possibilities in the context of gambling but Boltzmann could have also resorted to the posthumous paper [31] by the Reverend Bayes who had considered circumstantial possibilities in the context of collecting information [32]. It turns out that new insight to behavior and the payoff function can be obtained from a re-examination of the probability concept that Boltzmann placed as the cornerstone of his statistical mechanics.

Boltzmann enumerated, just like counting pips on dice, the isoenergetic configurations that are commonly referred to as microstates. This invariant probability notion, here referred to as Cartesian, is constant in energy and thereby it corresponds to stationary systems. Hence the statistical theory, by founding solely upon this, is limited to changes in configurations of conserved systems. In contrast, the Bayesian probability $P$ can be seen to vary due to changes in energetic conditions, and thereby it relates to evolutionary systems. Hence the statistical theory, based on the conditional probability notion, describes state changes of non-conserved systems. The equation of motion for an evolving system expresses the principle of increasing entropy

$$d_t S = k_B d_t \ln P = -\sum_{j,k} d_t N_j \, A_{jk} \big/ T = k_B L \geq 0 \,. \tag{1}$$

In the context of game theory the rate of entropy increase $d_t S$ is the payoff function that values the outcomes of various $jk$-transactions where $k$-assets of players are transformed to $j$-assets of other players (and vice versa). For example, a player uses money in his possession to buy goods from another. The actions bring about changes $d_t N_j$ in the $j$-assets of the player concurrently with changes $d_t N_k$ in the $k$-assets of the other player. In fact not only money and goods are rated by energy differentials but literally everything is valued in terms of energy differentials. Therefore the holistic formalism is able to describe the behavior of complex systems just as of simple systems. The quest to consume free energy in least time is ubiquitous and independent of mechanisms. The overall sequence of transactions, referred to as the course of a game, advances move-by-move when the free energy $A_{jk}$ = $\Delta\mu_{jk} - i\Delta Q_{jk}$ is consumed. The term $\Delta\mu_{jk} = \mu_j - \Sigma\mu_k$ contains the scalar potential differences in the assets where $\mu_j = k_B T \ln[N_j \exp(G_j/k_B T)]$ is a function of the number $N_j$ and value $G_j$ of the $j$-asset. When the driving force contains only the potential differences, the system is closed to net flows of energy from its surroundings. In the corresponding zero-sum game the transactions between players bring about merely an exchange of assets but the total status over all players remains invariant. The invariant nature of zero-sum games is stated by the minimax theorem [33]. The term $i\Delta Q_{jk}$ in Eq. 1 is the energy influx from the surroundings to an open, extensive-form game. This term imposed by the surroundings distinguishes the statistical physics of open systems from the conventional formalism used by Boltzmann and others ever since. The influx, which is often equated with income, is incorporated in the assets by the players' actions. Conversely, a player may lose his assets to the others as well as to the surroundings by misfortunate moves.

A move by a player will cause a change in assets. The rate of change is proportional to the driving force $A_{jk}$ by the coefficient $\sigma_{jk}$ [13]

$$d_t N_j = -\sum_k \sigma_{jk} A_{jk} / k_B T .$$

(2)

When $A_{jk} < 0$, the $jk$-transaction will increase $d_t N_j > 0$, and vice versa. The particular functional form in Eq. 2 ensures that conservation of energy is satisfied in every move [12]. When Eq. 2 is inserted in Eq. 1, entropy is found to increase almost everywhere $d_t S \geq 0$ because each square $A_{jk}{}^2 \geq 0$. This is the principle of increasing entropy.

The proportionality coefficient $\sigma_{jk}$ represents a particular mechanism that channels the $jk$-transaction. For example, a more effective means of trading brings about faster changes in the assets. The rate is not immaterial because the driving force ($A_{jk}$) is, in turn, a function of assets ($\mu_j$). In other words, behavior and the motives of behavior are inseparable from each other. Indeed, it is witnessed that when stakes are raised, behavior will change. For example, in the well known ultimatum game the probability of rejecting a share, whether fair or unfair, decreases with the increase of the absolute amount offered [34]. However, in classical game theory the amount of assets in players' possession is customarily ignored.

According to the physical portrayal of games more effective mechanisms are favored by the transactions themselves as they allow for a faster maximization of entropy. The quest to increase entropy in the least time is also known by the maximum entropy production principle [35]. According to the adopted self-similar thermodynamic formalism a game itself is also a mechanism that may evolve further to facilitate the overall consumption of free energy. In other words there are games being played within games, in accordance with hierarchical system theory [36]. For example, the behavior of a citizen amongst others can be regarded as a game that ultimately also contributes to international relations that, in turn, can be regarded as a game among nations. In the quest for the maximal dispersal of energy the systems will form a coalition, i.e., a larger system where it is more effective in acquiring and consuming free energy and distributing the acquired flows among its constituent systems [37].

Moreover, the functional form of $d_tS$ in Eq. 1 implies that the particular possessions of a player affect not only their decisions but those of others as well. The product form of Eq. 1 states that actions ($d_tN$) are inseparable from possessions ($A_{jk}$). This interdependency is the source of unpredictability when an open, extensive-form game is played by three or more parties [38]. The courses of non-deterministic games vary and they do not necessarily end up with the same outcome because a move at any stage depends on past moves and conversely restricts the future choice of moves (Fig. 1). At the branching points, to be precise, the derivate $d_tS$ (Eq. 1) is inexact. In terms of physics the game is a non-Hamiltonian system with three or more degrees of freedom where the driving forces and energy flows are inseparable [12,39]. This thermodynamic interdependency between flows of energy and the free energy that drives all other flows underlies the interdependency between the strategy $s_i$ and its complement, $s_{-i}$, the strategies played by all other players. In other words, the thermodynamic theory gives the reason why the decision made by a player is dependent on the decisions of other players, which, in turn, are dependent on the first player's decision. In fact repeatedly changing conditions may drive repeated changes of strategies. For example, it has been shown that in a well studied game, the iterated continuous prisoner's dilemma, no strategy is evolutionarily stable [40]. Moreover, the intractability of extensive-form games is understood as an inherent characteristic of open systems. Owing to the net influx or efflux of energy to or from the system, there is no norm and hence no unitary transformation either to obtain a solution or to predict the trajectory toward the solution concept. Even a minor move will perturb the energy content of the system and hence affect the future course. This is characteristic of chaotic games [41]. The sequence of moves, i.e., kinetics is, instead of using Eq. 2, often modeled by the deterministic law of mass action [42], but then thermodynamics and kinetics become incompatible with each other [12]. Alternatively, a sequence of moves is modeled as a Markov chain [43]. However, even when present probabilities depend on the past, the probabilities are not understood as physical.

At any given time the game, described as a natural process, can be assigned with the additive logarithmic status measure known as entropy [12]

$$S = k_B \ln P = k_B \ln \prod_j P_j \approx k_B \sum_j N_j \left( 1 - \sum_k A_{jk} / k_B T \right) \qquad (3)$$

over the product of probabilities $P_j$. Entropy will increase toward the steady state value $S_{max}$ where all free energy terms $A_{jk} = 0$. In this case no strategy, i.e., no choice of $jk$-moves can be found by any player to improve the distribution of assets $N_j$ among the players or to acquire more assets from the surroundings. Customarily, the optimal behavior is referred to as the evolutionary stable strategy (ESS). The free energy minimum state can be proven stable against perturbations $\delta N_j$ using Eqs. 1 and 3 in the Lyapunov criteria $S(\delta N_j) < 0$ and $d_tS(\delta N_j)/dt > 0$ [44,45]. In other words, there is no action that could improve the status and no strategy that any player could play that would return a higher payoff, given all the strategies played by the other players.

The steady-state partition of assets is given by the condition $d_tS = 0$ which yields from Eq. 1

$$N_j = \prod_k N_k e^{-\left(\Delta G_{jk} - i\Delta Q_{jk}\right)/k_B T} \tag{4}$$

This is the condition of reaction equilibrium [26]. The dispersal of assets at the free energy minimum is a skewed, nearly log-normal distribution [14]. Indeed, studies of behavior in many-players games reveal that the strategies of individuals will evolve and the game will end up in a stationary state where only few have gathered large assets while most players must have settled for moderate possessions and relatively few have only little. Both the maximum and the amount of skew for a specific situation depend on the overall energy content of the system (Fig. 1). This is familiar from the distributions of ecosystems [46] and economic systems [47] and from partitions of elementary chemical and physical systems [48].

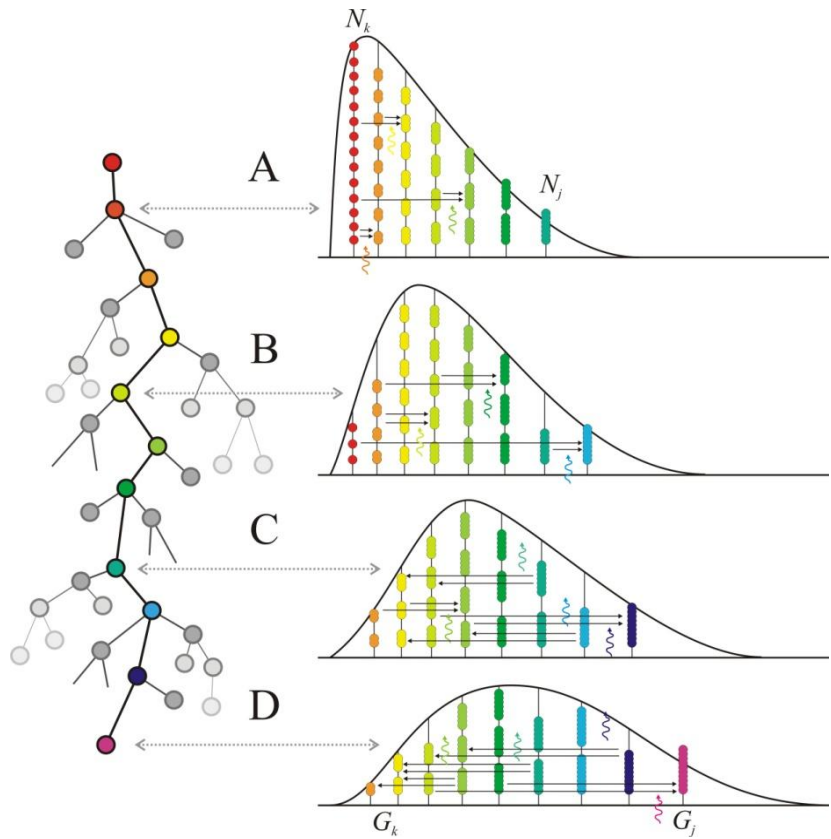

**Figure 1.** Extensive-form game branching on the left side is depicted on the right side as a natural process that brings about changes in diverse assets, indicated by the solids, that correspond to populations $N_j$ on energy levels $G_j$. As denoted, move by move, by arrows in the sequence of state changes (A – D) the payoff function as entropy increases until a solution concept corresponding to a stationary state of free energy minimum has been attained. This maximum entropy partition as a thermodynamic steady state is an evolutionary stable solution concept where the net dissipation (wavy arrows) vanishes. No move is able to bring more assets, i.e., energy to the process that settles to the evolutionary stable state.

The optimal distribution given by Eq. 4 is the result of a natural process just as it is the outcome of an extensive-form game. However, it is worth emphasizing that at the steady state there are no net fluxes to or from the systems just as the players' total assets at a particular solution concept are no longer changing. This parallels the fixed-energy condition of stationary systems. It means that even though the outcomes of repeated games differ, each of them is a feasible solution concept, i.e., the minimax condition [1] just as in a zero-sum game is satisfied for all players. Also, the proof of the folk theorem [2] rests on the invariant nature of an outcome [49].

## 4. Motives of Behavior

Is it reasonable, as argued above, to equate incentives of behavior with the imperative to increase entropy? Would not the association of animate actions with the physical principle of decreasing free energy quench all degrees of freedom from behavior? These and other concerns about the use of rate of entropy increase as the universal payoff function deserve to be addressed.

According to the naturalistic tenet, behavior is, just like any other natural process, limited by the free energy because a more generous stance would violate the conservation of energy. In other words, the free energy sums up all resources that an individual or any other agent has in possession or access to with its available mechanisms. Beyond that no one can act. This superior role of free energy is apparent, for example, from actions that a social system is able to take when fostering its subordinate individuals whereas an individual, even a rich one, rarely has enough power to act against society for any significant periods of time. The thermodynamic description of behavior as an energy transduction process leaves all of the available pathways open for conduction, but there is the natural bias for the best path, known also as a geodesic, that will diminish free energy as soon as possible [50]. Staying on an optimal trajectory requires both incessant evaluation of alternative pathways and redirection after each choice. In other words the game is changing as it is played. Thus behavior cannot be reduced to precisely predictable sequence of acts.

The rate of entropy increase in the least time (Eq. 1) as the motive of behavior measures all available sources of free energy ($A_{jk}$) weighted by rates of their consumption ($d_t N_j$). In this sense entropy, just like utility, considers numerous variables that influence behavior. However, since the payoff function of the rate of entropy increase evaluates everything in terms of energy, it makes all motives of behavior commensurable with each other. In practice, though, it may be difficult to accurately assign an energetic value to every option, but in a statistical sense the behavior itself reveals the current value of a strategy. Nonetheless, it is legitimate to question whether a single number can possibly sum up all, at times even conflicting, motives of behavior and eventually to anticipate behavior. The answer is yes only in the statistical sense. The statistical notion is applicable when any one move will not change the course of game too much. In physical terms, the condition is valid when the consumption of $A_{jk}$ is small in comparison to the average energy content in the system $k_B T$. However, events where $A_{jk} \approx k_B T$, may be infrequent, but they do happen. This long-tail trait of a probability distribution is characteristic of natural processes [51]. When $S = k_B \ln P$ (Eq. 3) is not a sufficient statistic for $k_B T$, the state of a game is given best by the probability [12,39]

$$P = \prod_j P_j = \prod_j \left( \prod_k N_k e^{-\left(\Delta G_{jk} - i\Delta Q_{jk}\right)/k_B T} \right)^{N_j} \Bigg/ N_j! . \qquad (5)$$

The exponential form reveals that any one $P_j$, for which $A_{jk} = \Delta\mu_{jk} - i\Delta Q_{jk} \gtrsim k_B T$, contributes substantially to $P$. The $jk$-move corresponding to the consumption of $A_{jk}$ will change $P_j$ significantly, i.e., beyond the statistical approximation $\ln N_j! \approx N_j \ln N_j - N_j$ which is applicable to large, quasi-stationary populations. In this case the probability of a small system will not evolve smoothly but moves in steps according to $d_t P = LP$ (Eq. 1). In practice this means that the rate of entropy increase cannot be used to extrapolate a specific scenario. For example, when an individual labeled with $j$ happens to strike a particularly prosperous deal, his status measured by $P_j$ will step up abruptly. Nevertheless, the status $P$ of the overall course of the game, comprising many players, will remain sufficiently statistical and it will not be affected all that much by a single move but moves smoothly.

The description of a game as a natural process is holistic so that any $jk$-move will, in principle, affect the status of any other player. The physical probability in Eq. 5 defines, by $jk$-indexing, the interdependency among the densities-in-energy. A pair or a set of decisions are referred to as strategic complements when they are constructively reinforcing one another and as strategic substitutes when they are destructive in offsetting one another. The course of a game depends on coherent moves, which in terms of physics, means that the flows of energy interfere with each other when the affine, curved energy landscape is in evolution toward the stationary-state flatness [52,53]. Both sequential and simultaneous moves are accommodated in the formulation (Eq. 1) but since velocities of energy flows are limited, ultimately by the speed of light, only the sequential actions display causality [39] to affect subsequent decisions whereas simultaneous moves are independent, which is a familiar notion in sealed first-price auctions. Moreover, individual behavior via social interactions has been understood as the mechanism that bonds together the affine energy landscape and generates its evolution [54].

The physical probability that is conditioned in energetic terms (Eq. 5) clarifies also why mimicking (imitation) is often a successful strategy. *A priori*, i.e., at an initial state it may not be obvious which particular move will consume free energy in the least time but pioneers will search for paths, e.g., by trial and error. Initially the optimality of a path is less important because just about any strategy will produce entropy. Moreover since no experience, i.e. references, has been accumulated, the optimality cannot be assessed. Therefore a successor, when mimicking the established behavior, will follow, if not the best, at least a reasonable trail formed by the path breakers. Explorations are per definition suboptimal moves. This is consistent with rational ignorance [55] which states that the act of acquiring information on the best possible strategy or path may be too costly compared to expected and uncertain benefits to the player. The thermodynamic theory shows that the mere move to set a path will change the setting for subsequent moves as well [56]. The probabilities of future decisions are affected by past acts, in other words a specific state of a game depends on its history. The conditional interdependence among strategies is also familiar from cemented suboptimal standards. It is tedious to improve a widely adopted standard simply because initial payoff will suffer from the limited scope of applicability of the reform. Conversely, marginal benefits for conformists are easily available, whereas significant gains are in the sight of a rebel. The conventional way of thinking as an established means of energy dispersal is preferred rather than making much additional effort to explore beyond paradigms. Players are motivated to explore new strategies when they see a possibility for greater payoffs, i.e., a larger perceived gradient in free energy than is consumed by the study itself. New strategies may emerge from intentional manipulation or sporadic fluctuations also known as random variations. For the payoff, i.e. the entropy increase, it does not matter whether the move is intentional or accidental, since only the outcome is valued. The changes may be, for example, mutations in a biological context or reorganizations in the brain in a decision making context. When a game is maturing toward a solution concept, most new strategies are not so successful, but some may still tap into potentials better than competitors and gain ground.

## 5. Discussion

Game theory accounts for behavior in remarkably diverse circumstances, yet it is pertinent to ask what behavior actually is. Game theory rationalizes behavior by modeling it as a game that aims at maximizing payoff. However, the nature of a game as a process and its objective has remained obscure. Here games are described as natural processes that increase entropy in the least time. This tenet, while founded on the $2^{nd}$ law of thermodynamics as the universal principle, may at first appear superficial and deficient as if it was neglecting important factors such as the role and asymmetric distribution of information among players. However, the naturalistic view is holistic in relating everything to everything in terms of energy. As the flow of energy is the sole means of conveying information, this means that a piece of information is also an asset. This stance is valid because any form of information is bound to its physical representation [57,58,59], which, in turn, is subject to the laws of thermodynamics. Furthermore, the accumulation of information, e.g., a learning process itself, can be understood as a natural process of formation and changing of paths (geodesics) for flows of energy that represent information. Therefore, the value of information gained by behaving in a certain way, i.e., by playing a certain strategy which acquires information for the future accumulation of assets has to be accounted for. Therefore a change in strategy is a natural consequence of accumulating assets because the acquired assets open new opportunities for the reduction of free energy. The acquisition of information about other players' types, e.g., to account for a Bayesian game [60], is contained in the physical formulation of games as natural processes. This physical correspondence is also reflected in the purification theorem. It states that mixed strategy equilibria can be obtained as the limit of pure strategy equilibria from a perturbed game of incomplete information. Physically speaking, a mixed state can be constructed as the limit of pure states that are perturbed by energy in mutual interactions. Information is energy in interactions, when defined in thermodynamic [58] rather than mathematical [61] terms.

The intimate interdependency between behavior and its motives, that are, physically speaking, flows of energy and free energy, is apparent when acquired knowledge is used to anticipate moves by others. The extensive-form game where knowledge is accumulated is directed toward a self-confirming equilibrium [62] just as a natural process spontaneously progresses toward a stable state. An extensive series of repeated games will eventually reveal, in mutual transactions, all characteristics of all players. Thus all conceivable paths of actions are open to maximize the total payoff. This revelation shifts the steady state from the Bayes-Nash equilibrium to the ultimate optimum [63,64]. This tedious optimization procedure relates to computational complexity [65]. There is no algorithm that would solve this non-deterministic polynomial time problem [66]. At the Lyapunov-stable state there are no unexposed assets and no strategies that could possibly shift the equilibrium any further.

The extensive-form game models the fact that behavior depends on surrounding conditions. A change in the surroundings relates to changes in the values of a payoff matrix, and more efficient strategies to which higher payoffs are available may emerge. Any such change will bring instability to the game. The approach of the new equilibrium may even be chaotic because any early move will affect the set of moves available in the future. Moreover, when a particularly effective strategy is found and executed, it may consume assets more rapidly than they are replenished. Consequently, after a period of overexploitation, the system must retract and abandon the once so lucrative, but unsustainable strategy. For this reason animal populations oscillate and economic cycles follow one another.

The connection between the entropy maximum and game theoretical equilibrium is, as such, not a novel proposition [67]. However, here the rate of entropy increase is provided in a mathematical form that is equivalent to the rate of free energy decrease. This is essential. Since everything can be valued in terms of energy, the rate of entropy increase qualifies as the universal payoff function. Moreover, statistical physics of open systems links the principle of increasing entropy to the principle of least action which guides processes along the optimal paths that bring the system to the stationary state in the least time. Admittedly, just like free energy, entropy in a complicated system is a function of many variables that are indexed by $j$ and $k$. In this way entropy as the additive figure of merit is of course very much in line with observations that human and other animate behavior is difficult, if not impossible, to account solely on the basis of a single motive. While justified by the profound principle it may be tedious to expand the entropy function in every detail to model practical situations. Rather it would seem sensible to model only the terms that are anticipated to be significant in decisions that are confronted in specified circumstances. However, thermodynamics clarifies that no precise predictions are available even from very detailed formulation because the moves themselves will change the conditions and prompt different moves. Moreover, the physical portrayal of games as natural processes by the rate of entropy increase as the payoff function gives justification for mixed and varying strategies over pure and fixed strategies. Diversity in behavior, like biodiversity, allows the entire system to consume free energy more and faster than would be possible via individual and invariant strategies.

Entropy as the payoff function also clarifies the subjective nature of decision making when choosing a strategy because payoff is a function of the possessions and moves that are available for a specific player. Moreover, in hopes of making a rational choice the player, who is equipped with appropriate knowledge, may discount a future payoff when making present-day decisions. Thus there is no universal rational choice, which is why some actions may seem irrational to an outside observer. In this sense the thermodynamic theory addresses some of the critical concerns about rational choice [68]. Furthermore, physical formalism does not allow for observations without interactions. Therefore an observer will inevitably affect, i.e., integrate himself in, the course of a game. Common values are approached via integration where superior free energy possessions and effective consumption strategies acquired by energy-intense players tend to impose on others what is deemed as rational, when in fact much of this "rationality" is in fact mimicry and submission to authority.

The physical portrayal of games as natural processes illuminates not only competition over assets but also cooperation among individuals. A coalition is regarded as a strategy, just as any other mechanism to increase entropy. The group possesses more means and more assets to access higher status in entropy than any one individual could master independently. The consensus in decision making is motivated only if it provides the means for each individual to attain a higher entropic status than would be available by independent moves. It is not unusual that, when circumstances change, coalitions will expire or reorganize to adapt to the new circumstances. This understanding adds to the ongoing debate concerning the emergence of cooperation, to which several solutions have been proposed [69,70]. Parameterization of models with a physical quantity may help to distinguish the type of game, for example snowdrift or prisoner's dilemma, to represent a situation.

Finally, the tragedy of the commons [71] that has also been analyzed by game theory [72] deserves clarification. The detrimental scenario that is driven by short-sighted individual incentives continues when resources and means of social bonding are insufficient. This alerting sequence of moves toward ruination is understood by the thermodynamic theory as probable. According to the natural law, when energy in the surrounding supplies falls, due to exploitation by individuals, below that contained in the social system, the flow of energy is redirected according to the $2^{nd}$ law away from the society to the surroundings. Consequently, the social system keeps draining its cohesion just when it would desperately need more energy to re-establish vital mechanisms such as social bonding to enforce co-operation that would be necessary for society to behave in a sustainable manner.

## Acknowledgements

We thank Niall Douglas, Vesa Kanniainen and Leena Kekäläinen for inspiring comments and valuable corrections.